\documentclass{article}

\usepackage[preprint]{neurips_2023}

\usepackage[utf8]{inputenc} 
\usepackage[T1]{fontenc}    
\usepackage{hyperref}       
\usepackage{url}            
\usepackage{booktabs}       
\usepackage{amsfonts}       
\usepackage{nicefrac}       
\usepackage{microtype}      
\usepackage{xcolor}         
\usepackage{tabularx}
\usepackage{booktabs}
\usepackage{graphicx}
\usepackage{enumitem}
\usepackage{cleveref}
\setlist[itemize]{leftmargin=*}
\setlist[enumerate]{leftmargin=*}
\usepackage{subcaption}

\title{LLM as OS, Agents as Apps: Envisioning AIOS, Agents and the AIOS-Agent Ecosystem}

\author{%
  Yingqiang Ge\\
  Rutgers University\\
  \And
  Yujie Ren\\
  Rutgers University\\
  \And
  Wenyue Hua\\
  Rutgers University\\
  \AND
  Shuyuan Xu\\
  Rutgers University\\
  \And
  Juntao Tan\\
  Rutgers University\\
  \And
  Yongfeng Zhang\thanks{\textbf{Author Affiliation}: Department of Computer Science, Rutgers University, New Brunswick, NJ, 08854, US; \textbf{Author Emails}: \{yingqiang.ge,yujie.ren,wenyue.hua,shuyuan.xu,juntao.tan,yongfeng.zhang\}@rutgers.edu}\\
  Rutgers University\\
}

\begin{document}

\maketitle

\begin{figure}[h]
\vspace{-20pt}
\centering
\includegraphics[width=0.8\textwidth]{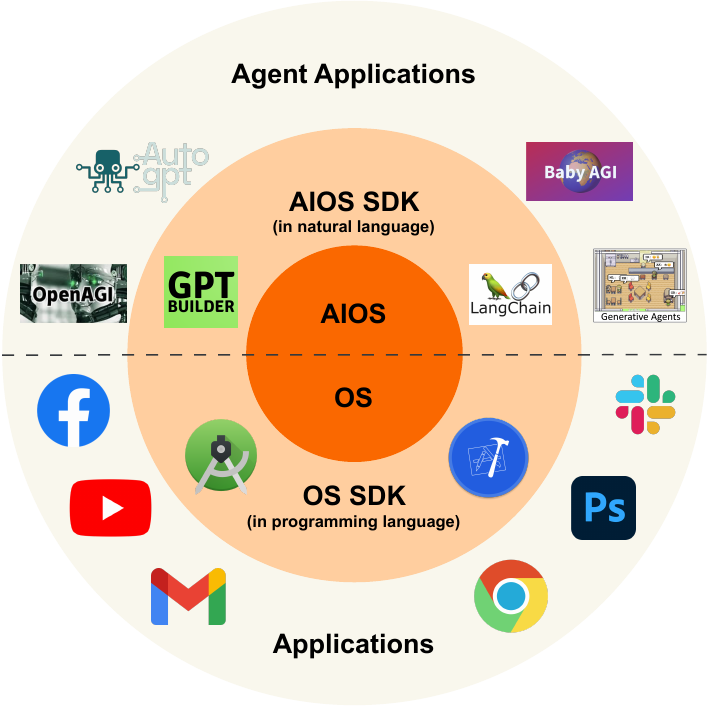}
\vspace{-5pt}
\caption{OS-APP ecosystem vs. AIOS-Agent ecosystem.}
\label{fig: ecosystem}
\end{figure}

\begin{abstract}



This paper envisions a revolutionary AIOS-Agent ecosystem, where Large Language Model (LLM) serves as the (Artificial) Intelligent Operating System (IOS, or AIOS)--an operating system ``with soul''. Upon this foundation, a diverse range of LLM-based AI Agent Applications (Agents, or AAPs) are developed, enriching the AIOS-Agent ecosystem and signaling a paradigm shift from the traditional OS-APP ecosystem.
We envision that LLM's impact will not be limited to the AI application level, instead, it will in turn revolutionize the design and implementation of computer system, architecture, software, and programming language, featured by several main concepts: LLM as OS (system-level), Agents as Applications (application-level), Natural Language as Programming Interface (user-level), and Tools as Devices/Libraries (hardware/middleware-level).

In this paper, we begin by introducing the architecture and historical evolution of traditional Operating Systems (OS). 
Then we formalize a conceptual framework for AIOS through ``LLM as OS (LLMOS),''\footnote{For convenience, LLMOS may be pronounced as ``el-mos''.} drawing analogies between AIOS components and traditional OS elements: LLM is likened to OS kernel, context window to memory, external storage to file system, hardware tools to peripheral devices, software tools to programming libraries, and user prompts to user commands.
Subsequently, we introduce the new AIOS-Agent Ecosystem, where users and developers can easily program Agent Applications (AAPs) using natural language, democratizing the development of and the access to computer software, which is different from the traditional OS-APP ecosystem, where desktop or mobile applications (APPs) have to be programmed by well-trained software developers using professional programming languages.
Following this, we explore the diverse scope of
Agent Applications. These agents can autonomously 
perform diverse tasks, showcasing intelligent task-solving ability in various scenarios.
We delve into both single agent systems and multi-agent systems, as well as human-agent interaction.
Lastly, we posit that the AIOS-Agent ecosystem can gain invaluable insights from the development trajectory of the traditional OS-APP ecosystem. Drawing on these insights, we propose a strategic roadmap for the evolution of the AIOS-Agent ecosystem. This roadmap is designed to guide the future research and development, suggesting systematic progresses of AIOS and its Agent applications.

\end{abstract}





\maketitle

\tableofcontents

\newpage


\section{Introduction}

In the evolving landscape of information technology, Operating Systems (OS) such as Windows\footnote{\url{https://www.microsoft.com/en-us/windows/}}, MacOS\footnote{\url{https://www.apple.com/macos/}}, iOS\footnote{\url{https://www.apple.com/ios/}}, and Android\footnote{\url{https://www.android.com/}} have become cornerstones of our digital lives. 
On top of the operating systems, a diverse range of applications (APPs) are developed, helping with users' diverse tasks and enriching the OS-APP ecosystem.
For example, Microsoft Word\footnote{\url{https://www.microsoft.com/en-us/microsoft-365/word/}} and Google Docs\footnote{\url{https://www.google.com/docs/about/}} excel in drafting documents, while Microsoft Outlook\footnote{\url{https://www.microsoft.com/en-us/microsoft-365/outlook/}} and Gmail\footnote{\url{https://www.google.com/gmail/about/}} offer efficient email management. 

Operating systems have advanced significantly, becoming more intuitive and user-friendly, yet their core remains rooted in static rules and predefined logic flows, without the intelligent, creative, and emergent task-solving abilities. The applications built on top of such OS, on the other hand, 
are also limited to their designed purposes, unable to transcend beyond their individual scopes. Whenever individual applications need to incorporate intelligent abilities, they have to implement their own AI methods or functionalities, sometimes based on third-party libraries.
This isolated framework underscores a significant shortfall in the current OS-APP ecosystem and highlights the pressing need of infusing (artificial) intelligence into operating systems, so that intelligence can be natively distributed to the various applications built on top of it.

As a result, this paper envisions (Artificial) Intelligent Operating System (IOS, or AIOS), an operating system ``with soul''. Furthermore, a diverse scope of intelligent Agent applications are built on top of the AIOS, leading to the new AIOS-Agent ecosystem, in comparison to the traditional OS-APP ecosystem, as shown in Figure \ref{fig: ecosystem}.
Due to the versatile and remarkable capabilities they demonstrate, Large Language Models (LLMs) \citep{radford2019language, brown2020language,touvron2023llama-2,alpaca} are regarded as potential sparks for Artificial General Intelligence (AGI) \citep{bubeck2023sparks,morris2023levels,ge2023openagi}, offering hope as foundational elements for the development of AIOS.
There are several reasons confirming LLMs' general capability and feasibility for building AIOS:
\begin{itemize}
    \item First, LLMs have demonstrated exceptional language understanding abilities as well as reasoning/planning abilities to solve complex tasks, which can divide the tasks into several sub-tasks and conquer them one-by-one, sometimes with the assistance of external tools \citep{ge2023openagi, wei2023chainofthought, huang2022towards}. 

    \item Second, LLMs offer a highly flexible platform to process virtually any prompt, instruction, or query expressed in natural language, making it possible for a diverse range of Software Development Kits (SDKs) and/or applications to be built on top of them.

    \item Third, LLMs offer a more intuitive and user-friendly interface, since they can understand and respond to user prompts or instructions in natural language \citep{brown2020language,touvron2023llama-2}. This sheds light on the future where natural language serves as the programming language, making technology more accessible, especially for those who may not be familiar with traditional computer interfaces and programming languages. 

    \item Fourth, LLMs can be programmed to learn from interactions and customize their responses based on user preferences and past interactions, providing a more personalized experience \citep{safdari2023personality,durmus2023towards}.
\end{itemize}

As a result, infusing intelligence into the OS-level through LLM makes it possible for easily distributing intelligent abilities into the application-level, providing a promising way to democratize intelligence across various applications.




\begin{figure}[h]
    \vspace{-10pt}
    \centering
    \begin{subfigure}{0.8\textwidth}
        \includegraphics[width=\textwidth]{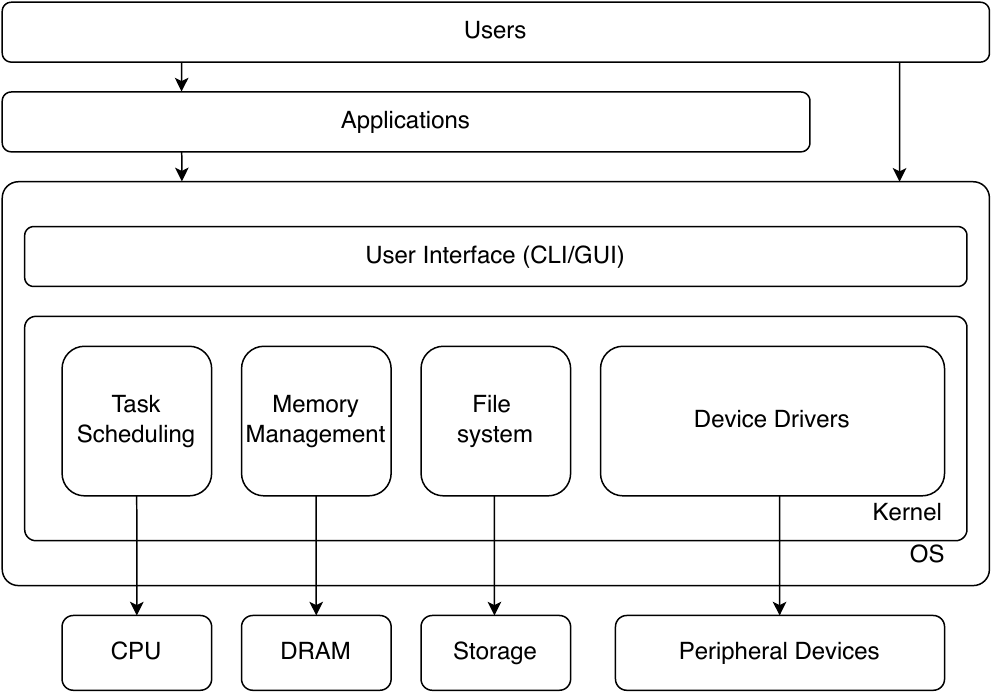}
        \caption{Architecture of Operating System (OS).}
        \label{fig: os}
        \vspace{5pt}
    \end{subfigure}
    
    \begin{subfigure}{0.8\textwidth}
        \includegraphics[width=\textwidth]{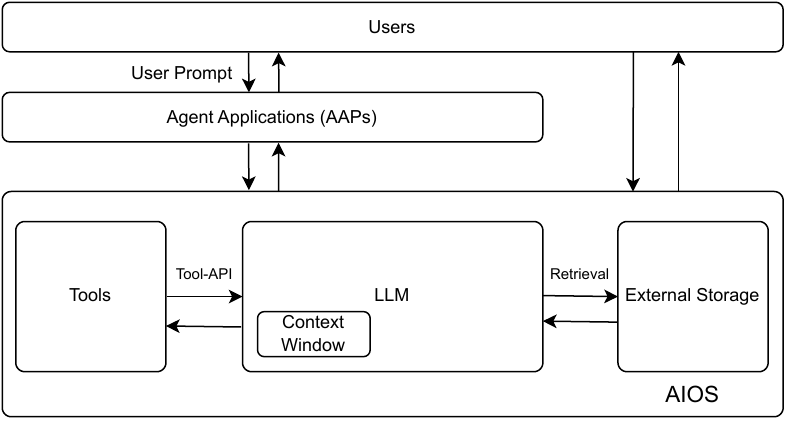}
        \caption{Architecture of Large Language Model as OS (LLMOS) for AIOS.}
        \label{fig: llmao}
    \end{subfigure}
\caption{Illustrations of the architectures of OS and AIOS (LLMOS).}
    \vspace{-20pt}
\end{figure}

Inspired by the architecture of traditional OS (as shown in Figure. \ref{fig: os}), we present a general framework for \textbf{LLM as OS (LLMOS)} with several key components in Section \ref{sec:llmao} (as shown in Figure. \ref{fig: llmao}): LLM as Kernel, Context Window as Memory, External Storage as File, Tools as Devices/Libraries, User Prompt/Instruction as User Interface (UI), and Agents as Applications.
The conceptual framework presented draws an analogy between an LLM as OS (LLMOS) and a traditional Operating System (OS), mapping various components of the LLMOS to elements of an OS. The LLM itself is likened to the kernel, the central part managing the system's core functions. The LLM's context window is compared to the memory of an OS, handling immediate context selection and data processing \citep{shi2023large}. External storage for the LLM is analogous to the files of an OS, allowing for long-term data storage. Meanwhile, there are corresponding data retrieval methods to enable retrieval-augmented LLMs \citep{guu2020retrieval}, which serve as the file system of OS to manage and find relevant files. Besides, LLM can make use of various tools for task solving \citep{ge2023openagi}, including both hardware tools and software tools. Hardware tools in the LLMOS framework are equated to peripheral devices in a traditional OS, each offering specific functionalities to help interact with the physical world, while software tools in the LLMOS framework serve as the programming libraries in traditional OS, enabling Agent applications to interact with the virtual/digital world. User prompts or instructions for the LLM are akin to the user interface (UI) in a traditional OS, facilitating interaction between the user and the system. The user prompts or instructions can be direct natural language instructions provided by the user, and they may also be (sometimes semi-structured) natural language instructions converted from users' non-natural language instructions such as clicking on icons.
This mapping provides a systematic way to understand the operational similarities between LLMOS-based AIOS and traditional OS.

Upon establishing a robust conceptual framework for LLMOS-based AIOS, we introduce the AIOS-Agent Ecosystem, akin to the traditional OS-APP Ecosystem, in Section \ref{sec:ecosystem}. We begin by introducing the concept of Agent Applications (AAPs) within LLMOS, analogous to traditional applications (APPs) based on an operating system. These AAPs represent a diverse scope of specialized tasks for users to execute based on LLMOS. As illustrated in Figure \ref{fig: agent}, by integrating the LLMOS layer with the OS layer, Hardware layer, and the Agent Application layer, we can construct an autonomous AI Agent system. Such AI Agent system responds to user prompts or instructions in natural language and is capable of performing a multitude of tasks based on its interaction with the physical or digital environment. Since a diverse scope of Agents can be developed on top of the shared AIOS foundation, this eventually leads to an AIOS-Agent ecosystem. Moreover, in the new AIOS-Agent Ecosystem, the users and developers can easily program Agent Applications (AAPs) using natural language, democratizing the development of and the access to computer software, which is different from the traditional OS-APP ecosystem, where desktop or mobile applications (APPs) have to be programmed by well-trained software developers using professional programming languages.

Following this, we further delve into the practical LLMOS-based Agent Applications in Section \ref{sec:agents}. This section explores the potential of enhancing LLMOS's functionality by developing various agents in the real world, and further investigates the dynamic interactions between multiple agents and humans.
LLMOS-based agents are characterized by their creative autonomy, enabling them to generate novel ideas, narratives, or solutions not pre-programmed into them \citep{Chase_LangChain_2022,Significant_Gravitas_Auto-GPT,ge2023openagi,li2023camel,yao2022react,yao2022webshop}, which is indicative of an advanced level of creative intelligence \citep{yuan2022wordcraft,franceschelli2023creativity}. Specifically, Section \ref{sec:single} discusses applications of single agents, Section \ref{sec:multi-agent} explores multi-agent systems, and Section \ref{sec:human-agent} focuses on human-agent interactions.

Finally, in Section \ref{sec:future}, we explore several crucial future research directions for LLMOS and AIOS in general, drawing parallels and learning from the evolution of traditional operating systems. These directions span a wide array of areas, aiming to enhance the capabilities and applications of LLMOS: 1) \textbf{Resource Management.} For example, OS employs virtual and shared memories to address the issue of limited physical memory. LLMOS can inspire from these ideas to mitigate its own problem of limited context window challenges; 2) \textbf{Communication.} Different OSes and applications communicate using standardized protocols (such as Domain-Specific Languages); LLMOSes and Agents can build and use similar standard protocols for exchanging data and instructions with various systems, ensuring compatibility and smooth interaction across diverse platforms; 
3) \textbf{Security.} 
Security vulnerabilities in OS are important issues. State-of-the-art approaches aim to detect and capture malware and viruses at various levels. Similarly, LLMOS can implement detection and intervention mechanisms to regulate and monitor the implementation of third-party tools and Agent Applications.

\section{Aligning LLM and OS}
\label{sec:background}
\subsection{OS and Connections with LLM}
\label{bg:os}

The von Neaumann architecture, laying the foundation of modern computer hardware system, manipulate electrons and gates in the binary world, whereas human beings communicate with natural languages. Such gigantic semantic gaps between human users and computer hardware motivates an intermediary software layer interacting with users with a protected and abstracted view of underlying hardware resources such as CPU (Central Processing Unit), GPU (Graphics Processing Unit), RAM (Random Access Memory), storage and various other devices, which is called the Operating System (OS). The modern operating systems, over the past few decades, have evolved in a multi-layered architecture with modular components in each layer. This design not only enhances the efficiency and functionality of the systems but also facilitates easier management, scalability, and integration of diverse hardware and software elements.

\subsubsection{Kernel}
\label{bg:os:kernel}

The kernel, as its name suggests, encapsulates a set of core functionalities of managing hardware resources (e.g., CPU, GPU, DRAM, storage, and devices) as a nutshell.\footnote{We limit our scope to traditional monolithic kernel design.}

\begin{itemize}
\item \textbf{CPU Management} (Process/Thread, scheduling). To manage the CPU resources, modern operating systems abstract the execution of user programs or applications on a physical CPU as processes or threads. When the user launches an application, the operating system kernel loads the executable binary files of this application or program into DRAM, create the necessary data structures to book-keep any running states for this program, and allocate necessary resources.

However, the power wall~\citep{power-wall} limits the number of physical CPUs to be integrated into a single chip. To provide users with the illusion that they possess physical CPUs without sharing with others, modern operating systems multiplex the running processes or threads with limited number of CPUs with time-sharing~\citep{UNIX-time-sharing} and more dedicated policies~\citep{Linux-CFS, EEVDF-stoica}, much like multiple users share the same LLM backbone when processing their prompts or instructions. In such cases, if the inputs from multiple users are beyond the capacity of the resources, such as the tools, then the LLM may perform scheduling of the user prompts.


\item \textbf{Memory Management.} The physical memory in a computer, also known as DRAM (Dynamic Random Access Memory), is the key component to store the instructions and data of both the OS and applications. The OS is responsible for managing and allocating free space in the physical memory from application's requests. 

As described above, the physical memory, depicted as ``dynamic,'' cannot store data persistently when power is off, as it needs to refresh periodically to prevent the loss of data. In addition to its volatile nature, the evolution of memory falls behind the CPU for a long time, which is known for ``The memory wall''. This fact lays in two orthogonal aspects. First, the data transfer rate between DRAM and CPU can no longer catch up with the speed that CPU processes data. Second, the memory capacity on a single node stops scaling. Although the emergence of Compute eXpress Link \citep{CXL} alleviates the memory capacity wall, it still cannot keep the rapid pace of data growth at the artificial intelligence era. 

This is much like the context window of LLMs, which is usually limited by the maximum number of tokens that the LLM can handle. Besides, LLM usually needs to select the relevant information from the input context, since not all contexts are relevant for the current task and LLM may be easily distracted by irrelevant contexts \citep{shi2023large}. Context selection can be either implicitly realized through attention mechanisms, or be explicitly implemented by selecting relevant segments from the input context, much like the memory management process by traditional OS.

\item \textbf{Storage Management.} Storage devices, which store data persistently, provide much more density than memory with fewer cost, but much slower. The operating systems abstract the storage as file, and organize files in a systematic way with a component called the ``file system''. The file system contains the meta-data to index the actual data stored on the storage media. 

In addition to the file abstract, modern operating systems reserve a small portion of storage as an extension of memory, which is called the ``swap area''. It is the operating system that tracks the hotness and coldness of the data from user applications, and swaps code data from physical memory to the swap area on the storage devices.

Similarly, LLM often has access to external data storage for retrieval-augmented language modeling \citep{guu2020retrieval}. These external data can be free text data, structured tabular data, semi-structured graph data, or others. Furthermore, the external data are usually properly indexed for efficient and accurate retrieval, much like the storage management process of traditional operating systems.

\item \textbf{Device Management.} Peripheral devices, often excluded from the core computing system of CPU and DRAM, form an important set of functionalities for user input and output. Those devices range from mouse, keyboard, to GPU and network interface cards (NIC). The operating systems take the responsibility to manage those devices to orchestrate with other core components in the OS kernel.

There are thousands of different devices for different purposes and from different vendors all around the world. Hence, it is not possible to implement the driver program for all of those devices. Instead, modern operating systems expose a generic interface as device driver APIs for device vendors, and thus shifts the responsibility of developing device driver programs from OS maintainers to vendors of those devices. To provide a 'plug-and-play' feature, modern operating systems such as Linux include some universal and necessary device drivers for some common devices; for example, the drivers for GPU and USB devices~\citep{LDDv3}.

Similarly, large language models are not only text-in-text-out models, instead, they have the ability of leveraging various tools for solving complex tasks \citep{schick2023toolformer, ge2023openagi}. There can be two types of tools, hardware tools and software tools, which help the LLM to interact with the physical world and the digital world, respectively. In this context, the hardware tools for LLM are similar to the devices for traditional OS. Furthermore, just like driver programs connect devices with OS, Tool-Drivers can be developed to connect LLM and hardware tools, so that LLM can easily leverage the tools for task-solving. We will discuss software tools in the following.

\item \textbf{SDK and Programming Libraries.} The SDK (Software Development Kit) and programming libraries of an operating system are crucial tools that enable developers to easily create applications. 
They serve as the backbone of application development for operating systems, not only enabling and simplifying the creation of applications, but also ensuring that these applications are secure, efficient and compatible with the OS, which significantly contribute to the vitality and growth of the OS-APP ecosystem.

Similarly, LLM and the various AI Agents built on top of it can make use of various software tools such as searching on the web, checking for weather conditions, and booking for flight tickets \citep{ge2023openagi}. These software tools serve as reusable functionalities that can be leveraged by LLM and Agents for complex task-solving, and they can be provided as SDK or programming libraries so that users or developers can easily use them.

\end{itemize}

\subsubsection{User Interface}
\label{bg:os:ui}

As detailed previously, the kernel manages the hardware resources with proper abstraction for user to utilize. In order to enhance the accessibility of virtualized hardware resources, it is crucial to establish an interface between user and OS. 

\begin{itemize}
\item \textbf{System Call.} The system call, as the channel between OS kernel and users, defines a set of core functions to allocate and use the virtualized hardware resources. For instance, in a POSIX-compliant operating system~\citep{POSIX-standard}, the \texttt{mmap} system call allocates and manipulates memory resources. The \texttt{fork} and \texttt{exec} system call family deals with process and thread creation. The \texttt{read} and \texttt{write} system call are used to interact with storage devices. In terms of LLMs, the system calls can be formulated as natural language prompts into instruct the LLM for task execution. 

\item \textbf{Command-line Interface (CLI).} The command-line interface defines a set of utility programs built on top of system calls to facilitate users to operate on the computer hardware in an interactive manner. Users interact with the OS in those utility programs as commands. For instance, the \texttt{cd} and \texttt{ls} commands implement the action of entering a directory and list all the files and folders in a directory. In the context of LLM, natural language prompts naturally serve as the interface for users to interact with LLMs. Furthermore, the LLM may also pre-define some foundational and commonly used functionalities as standard prompt templates for users to use, similar to the standard commands as in traditional OS.

\item \textbf{Graphic User Interface (GUI).} The graphic user interface is a visual way for users to interact with computers and electronic devices using graphical elements such as icons, buttons, windows, and menus, as opposed to a text-based interface such as a command-line interface. GUIs make it easier for users to interact with complex systems by representing actions through visual elements and providing a more intuitive user experience, especially with the increasing demand and use of mobile devices such as smart phones discussed in~\cref{bg:os:history}. In terms of LLMs, graphic user interfaces can also be developed for LLM and Agents so that users can more conveniently interact with them without the need to writing long prompts. Instead, these GUIs will convert user's non-language instructions (such as clicking on icons) into (sometimes semi-structured) natural language prompts based on pre-defined prompt templates, and these converted natural language prompts will be sent to LLM for executing the user instruction.
\end{itemize}

\subsubsection{Operating System Ecosystem}
\label{bg:os:ecosystem}

The operating system ecosystem functions as an extension of the operating system, providing a comprehensive set of developer tool-kits (OS SDK) and runtime libraries, shown in Figure \ref{fig: ecosystem}. These tools empower application developers to efficiently design, implement, and run their applications within the operating system environment. For instance, the well-known iOS ecosystem includes a dedicated application development toolkit known as Xcode\footnote{\url{https://developer.apple.com/xcode/}}, alongside an application publishing platform called the AppStore\footnote{\url{https://www.apple.com/app-store/}}, complementing the core iOS ecosystem. In this ecosystem, the OS provides a bunch of resources to support APP development and also services as the platform for deploying and hosting these APPs, which eventually leads to a prospering OS-APP ecosystem.

Similarly, we envision an AIOS-Agent ecosystem, where LLM serves as the operating system and hosts a diverse range of AI Agent applications, as shown in Figure \ref{fig: ecosystem}. The LLM as OS (LLMOS) environment shall also provide a comprehensive set of AIOS SDKs and/or libraries, predominately supporting programming in natural language, to help developers or even average users without any knowledge on professional programming languages, to easily develop and deploy Agent applications in the LLMOS-based AIOS-Agent ecosystem.





\subsubsection{Evolution History of Operating Systems}
\label{bg:os:history}

\begin{itemize}

\item \textbf{Batch Processing System.} Early batch processing systems were a fundamental aspect of the early days of computing, dating back to 1950s~\citep{OS-history}. These systems were characterized by a sequential execution of tasks, where jobs were submitted in batches for processing. Early batch processing systems laid the groundwork for subsequent developments in operating systems. While lacking the interactivity and responsiveness of modern systems, they played a crucial role in advancing computing capabilities and setting the stage for more interactive and user-friendly computing environments in the years to come.

\item \textbf{Time Sharing.} Time-sharing systems~\citep{OS-history}, as proposed in \textit{Multics}~\citep{MULTICS} represent a significant advancement in the history of operating systems, providing a departure from traditional batch processing systems and introducing the concept of shared, interactive computing. Many concepts introduced in time-sharing systems, such as multitasking, interactive interfaces, and dynamic resource allocation, have become integral parts of modern operating systems, which laid the groundwork for user-friendly computing environments, enabling efficient resource utilization and interactive computing.

\item \textbf{Multitasking.} As the hardware evolved to multiple cores, planning user tasks on available multi-core CPU is critical to maximize the CPU utiliztion. Multitasking involves scheduling processes to run on the CPU in a way that gives the appearance of concurrent execution. Process scheduling algorithms determine the order in which processes are executed. The UNIX operating system~\citep{UNIX-time-sharing}, developed in the late 1960s and early 1970s at Bell Labs, introduced the concept of processes, each with its own address space, and implemented a simple and efficient multitasking model.

\item \textbf{Visualization (GUI).} As described previously in~\cref{bg:os:ui}, the command-line interface used to be the narrow bridge between users to interact with OS. Since command lines are highly professional, it prevented broader groups of users from easily and efficiently operating the computers. From Xerox Alto introduced by Palo Alto Research Center (PARC) in 1973\footnote{\url{https://spectrum.ieee.org/xerox-alto}}, to Apple Macintosh introduced in 1984\footnote{\url{http://apple-history.com/128k}}, to Microsoft Windows\footnote{\url{https://winworldpc.com/product/windows-3/31}} introduced in the early 1990s, and to a wide range of open-source GUIs for Linux such as GNOME\footnote{\url{https://www.gnome.org/}}, KDE\footnote{\url{https://www.kde.org/}} and UNITY\footnote{\url{https://unityd.org/}}, the development of GUIs has significantly influenced the accessibility and usability of computers, making computing more intuitive and user-friendly.

\item \textbf{Cloud Computing.} Performing data computing and storage on a single computing node was no longer sufficient as the data scale grew significantly in the early 2010s. The development of client-server architecture in the early 1990s, where multiple clients connect to centralized servers, laid the foundation for distributed computing with the support of networked environments and the communication between clients and servers that emerged as OS core functionalities. 
More technologies such as virtualization and resource disaggregation empower the modern cloud computing community and market. 
Though cloud computing is not directly tied to the history of operating systems, the evolution of cloud computing has had a profound impact on how operating systems are designed, deployed, and utilized.

\item \textbf{Mobile and Embedded Systems.} Aligning with the Moore's law, the size and the computing power of a general-purpose CPU redefines the capability of modern embedded devices. Tailored for operating on a relatively constrained computing resource, modern mobile OS are specifically focused on power efficiency, network efficiency, optimized graphics for touch screen, and the prospering application ecosystem (detailed in~\cref{bg:os:ecosystem}). Similarly, embedded devices ranging from IoT to robotics pose the aspect of real-time responses and more strict resource management in OS, which has been reflected in many successful embedded OSes such as VxWorks\footnote{\url{https://www.windriver.com/products/vxworks}} and QNX\footnote{\url{https://blackberry.qnx.com/en}}. 

\item \textbf{AI-powered Features.} In the past decade, artificial intelligence (AI) technologies have experienced explosive growth. Specifically, breakthroughs in several areas of AI, such as natural language processing, computer vision, and speech recognition, are extending the interactive interface between users and OS to a new level. As examples of early implementations, Siri\footnote{\url{https://www.apple.com/siri/}} from Apple and Cortana\footnote{\url{https://www.microsoft.com/en-us/cortana}} from Microsoft, the AI-powered virtual assistants provide a rich set of capabilities to understand the requests from the users, such as message sending, call making, question answering, web search, recommendation, and controlling smart home devices. 

\end{itemize}


\subsection{AIOS, LLMOS and AI Agents}






Recent advances in foundation models, such as Large Language Models (LLMs), have significantly changed how AI applications are designed, trained, and used, including but not limited to NLP, CV, Search, RecSys, Multimedia, and Game applications. We envision that the influence of LLMs will not be limited to the application level, instead, it will in turn revolutionize the design and implementation of computer system, architecture, software, and the system-application ecosystem. This is realized through the following key concepts.

\subsubsection{LLM as OS (system-level)}
    LLM serves as the foundation AIOS platform that provides intelligent computing, APIs, and services that support various applications and manage various computing resources; Different from traditional OS, which aims at precision and efficiency, AIOS is characterized by its ``intelligence,'' which enables its intelligent and creative interaction with various applications and computing resources for task solving, leading to an operating system ``with soul''.

\subsubsection{Agents as Applications (application-level)}
    Based on the LLM-driven AIOS, various AI Agents can be developed as AIOS-native applications, such as trip planning agent, medical consulting agent, financial management agent, etc.; these agents make use of the intelligent computing power of the LLM and the various tools provided by the AIOS SDK to solve various tasks; except for single-agent applications, agents may also communicate and interact with each other, enabling multi-agent applications; agent can even be created when needed and released after use, enabling on-the-fly creation of agents. There are several reasons that many specialized agents will be developed instead of integrating all of the functionality into one LLM: 1) the tasks that may be initiated by users are diverse, unbounded and unable to predetermine, 2) tasks may require specialized tools, reasoning and planning abilities, accessing the external physical or digital world, or domain-specific knowledge to complete, which the LLM-based AIOS platform may not support and need to be developed at the agent-level, 3) though language is a very powerful medium for describing tasks, objects, information or ideas, it may not be able to describe everything, and the completion of certain tasks may need creative forms of interaction beyond the ``language input, language output'' paradigm, such as the processing of vision, sound, touch, smell, and the diverse types of signals from various sensors. The processing of these unbounded and unpredictable types of information needs to be handled at the agent-level instead of the AIOS-level.

\subsubsection{Natural Language as Programming Interface (user-level)}
    In the AIOS-Agent ecosystem, average users can easily program Agent Applications (AAPs) using natural language, democratizing the development of and the access to computer software, which is very different from traditional OS-APP ecosystem, where desktop or mobile applications (APPs) have to be programmed by well-trained software developers using professional programming languages. This trend is consistent with the evolving history of programming languages, which are becoming more and more user-friendly, beginning from binary codes to assembly language to various high-level languages such as C, C++, Java and Python; Natural Language as Programming Language is a natural extension of this evolving history, making it possible for average users to program agent applications and interact with computers without special training on professional programming languages. 

\subsubsection{Tools as Devices/Libraries (hardware/middleware-level)}
    Just like traditional OS can take advantage of various input and output devices such as keyboards and printers to support various APPs for task fulfillment, AIOS can also provide various tools as services to support the agents' task fulfillment. Such tools include both hardware tools such as devices and software tools such as plugins or libraries;
    the tools also include both input tools such as collecting signals from a sensor and output tools such as sending messages to other agents;
    the AIOS shall include the basic and foundational tools that are frequently used by many agents as part of the platform, and provide Tool-Drivers (for hardware tools) and tool-APIs (for software tools) to enable easy calling of such tools by agents; Besides, each agent may also include its own specialized tools used for its own tasks.
    

\subsection{Development of OS and AIOS Aligned}

\begin{figure}[h]
\vspace{-5pt}
\centering
\includegraphics[width=\textwidth]{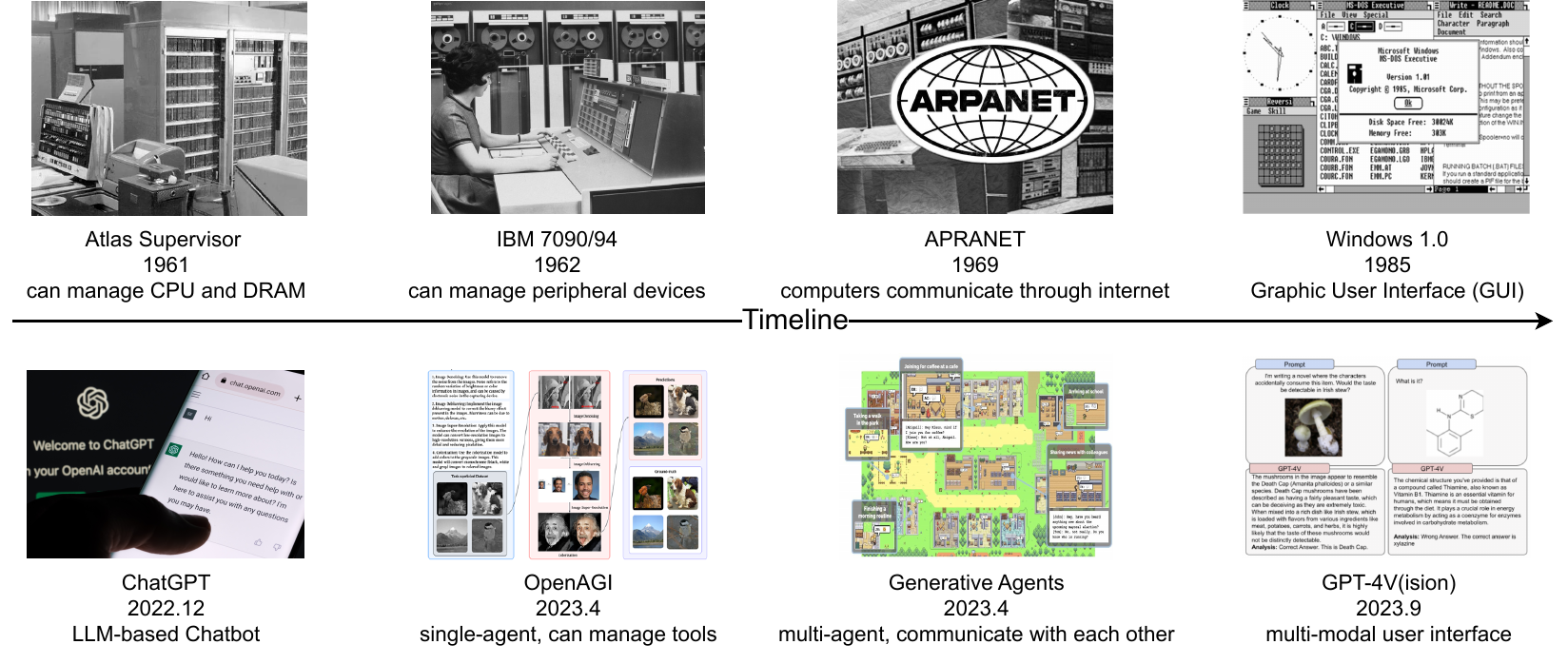}
\caption{Parallel Development of OS and AIOS/LLMOS.\protect\footnotemark}
\label{fig: parallel_development}
\end{figure}
\footnotetext{Image Sources:\\
\url{https://history-computer.com/atlas-computer/},\\
\url{https://en.wikipedia.org/wiki/History_of_Unix},\\
\url{https://www.magzter.com/stories/technology/Gadgets-Philippines/ARPANET},\\
\url{https://en.wikipedia.org/wiki/Windows_1.0x},\\
\url{https://www.brookings.edu/articles/chatgpt-educational-friend-or-foe/},\\
\url{https://github.com/agiresearch/OpenAGI},\\
\url{https://github.com/joonspk-research/generative_agents}, 
\\
\url{https://openai.com/research/gpt-4v-system-card}}

Figure \ref{fig: parallel_development} draws a parallel between the evolution of Operating Systems (OS) and LLM as OS (LLMOS). This comparison highlights their parallel progression in terms of enhanced functionality and user interaction. Initially, operating systems such as Atlas Supervisor were designed to manage basic computer resources such as CPU and DRAM. This evolved into more sophisticated version, exemplified by UNIX which can manage various peripheral devices. In a similar vein, LLMs have progressed from text-in-text-out chatbots to intricate LLM-based agents capable of managing various tools for complex task solving, as seen in OpenAGI \citep{ge2023openagi}. The figure also emphasizes the pivotal role of ARPANET in the development of TCP/IP protocols, which laid the foundation for today's Internet and connects multiple computers for communicating with each other. This is paralleled by the advancement of LLM-based multi-agent systems where agents can communicate with each other, signifying a trend towards more interconnected and collaborative LLM-based computing environments \citep{park2023generative}. Additionally, the evolution of operating systems is marked by the development of sophisticated graphical user interfaces (GUIs), such as those in Windows and Apple Macintosh. Similarly, LLM is evolving to include multi-modal interfaces, as demonstrated by GPT-4V(ision) \citep{openai2023gpt-4v}. This comparison underscores the role of both OS and LLM in revolutionizing user interaction with technology, transitioning from managing fundamental components to facilitating more intuitive, user-centric experiences.

\section{Architecture of AIOS} \label{sec:llmao}
In this section, we establish the conceptual framework of ``LLM as OS (LLMOS),'' creating parallels between LLMOS components and traditional OS elements, as is shown in Table \ref{tab:components}. In this analogy, the LLM is compared to the OS kernel, the context window to memory, and external storage to the file system. Tools and user prompts are equated with devices/libraries and the user interface, respectively. We will delve into the specifics of this correlation in the following discussion.

\begin{table}[h!]
\small
\centering
\begin{tabular}{>{\raggedright}p{2.6cm} >{\raggedright}p{3.2cm} p{6.6cm}}
\toprule
\textbf{OS-APP Ecosystem} & \textbf{AIOS-Agent Ecosystem} & \textbf{Explanation} \\
\midrule
Kernel & LLM & The core of AIOS, providing supportive services for Agent Applications (AAPs).\\
\addlinespace
\midrule
Memory & Context Window & Short-term memory of the current session, such as the prompt-response history of this session.\\
\addlinespace
\midrule
Memory Management & Context Selection and Management & Select relevant context for the session, and manage the context such as adding, deleting and changing context information. \\
\addlinespace
\midrule
File & External Storage & Long-term storage of the AIOS's history sessions, user profiles, factual knowledge, etc. \\
\addlinespace
\midrule
File System & Retrieval-augmentation & Retrieve relevant information from long-term storage. \\
\addlinespace
\midrule
Devices/Libraries & Hardware/Software Tools & Help systems interact with the external world (both physical and virtual/digital world). \\
\addlinespace
\midrule
Driver/API & Tool-Driver/Tool-API  & Serves as the interface for LLM/Agents to access and use the tools, usually in the form of prompts. \\
\addlinespace
\midrule
OS SDK & AIOS SDK  &  Assist users in developing Agent Applications. \\
\addlinespace
\midrule
User Interface (UI) & User Prompt/Instruction &  Instruction(s) given by user to the system in (sometimes semi-structured) natural language (NL) to perform specific tasks. The NL instruction may be converted from users' non-NL instructions such as clicks.\\
\addlinespace
\midrule
Application (APP) & Agent Application (AAP) &  A diverse world of AI Agents. \\
\addlinespace
\bottomrule
\end{tabular}
\vspace{5pt}
\caption{Comparison of OS-APP Ecosystem and AIOS-Agent Ecosystem}
\label{tab:components}
\end{table}

\subsection{LLM (as AIOS Kernel)}

The kernel is a set of computer programs at the core of a computer's operating system and generally has complete control over everything in the system. When a user command is issued, the OS parses the command and translates it into one or more system calls to enter the kernel, which are precise requests for the kernel to perform tasks such as process creation, memory allocation, and file manipulation. It then manages these tasks by scheduling the processes, allocating the necessary resources, interacting with device drivers for hardware access, and enforcing security measures. Throughout this process, the kernel also handles error checking and provides appropriate feedback. Upon completion, the kernel ensures that outputs are directed back to the user and that all resources are cleaned up to maintain system stability, effectively abstracting the intricate hardware details from the user.

Similarly, LLM acts as the operational core, akin to an OS kernel, responsible for planning and decomposing user requests, selectively calling tools, performing retrieval, and integrating all the information from previous steps to generate the final response. An LLM handles a complex prompt or instruction by first interpreting the input to understand its context and intent, and then uses its internal parameters to process the information, decomposing the instruction into manageable sub-tasks. The LLM generates responses for these sub-tasks, ensuring coherence and relevance to the original instruction. Finally, it synthesizes these into a cohesive output delivered to the user.

\subsubsection{Reasoning and Planning}\label{sec:planning}
In LLMOS, the role of LLM as the kernel subtly echoes the dynamics illustrated in the Dining Philosophers problem from computer science, a scenario that highlights the challenges of resource allocation and synchronization. In this problem, philosophers seated around a table must navigate shared resource usage--forks--in a way that avoids deadlock, a situation where no progress is made due to mutual blocking.
Reflecting these challenges, the kernel in an operating system is adept at managing and scheduling resources to avert system deadlocks, ensuring smooth and conflict-free operations. This essential function of the kernel in maintaining system stability and efficiency finds a parallel in the role of the LLM within the AIOS framework. Here, the LLM is responsible for navigating complex informational environments and managing external resources. 
Crucial to the LLM's role is its planning ability, which involves generating a series of actions to achieve a specific goal \citep{ge2023openagi}.
Central to the planning ability is the capability of reasoning \citep{bratman1988plans,russell1995prentice,fainstein2015readings,huang2022towards}. Through reasoning, LLMOS deconstructs complex tasks from user instructions into more manageable sub-tasks, devising appropriate plans for each. Next, we will explore several representative planning strategies that illustrate this capability.




\begin{itemize}
    \item \textbf{Single-path Planning.} In this strategy, the final task is decomposed into several intermediate steps, which are connected in a cascading manner, with each step leading to only one subsequent step. For example, CoT (Chain of Thoughts) prompting \citep{wei2023chainofthought}, as one of the celebrated capabilities of recent LLMs, is a pivotal breakthrough for performing complex multi-step reasoning when provided with limited examples. For example, by providing the models with ``chain of thoughts'', i.e., reasoning exemplars, or a simple prompt ``Let’s think step by step'', these models are able to answer questions with explicit reasoning steps \citep{kojima2022large}. ReAct (Reasoning + Acting) \citep{yao2022react} is a prompt-based paradigm to synergize reasoning and acting in language models. It enriches the action space with self-thinking (or thoughts), which compose useful information by reasoning over the current context to support future reasoning or acting. RAFA (Reason for Future, Act for Now) \citep{liu2023reason} studies how to complete a given task provably within a minimum number of interactions with the external environment.

    \item \textbf{Multi-path Planning.} In this strategy, the reasoning steps for generating the final plans are organized into a tree-like or graph-like structure. This mirrors the nature of human decision-making, where individuals often encounter a range of choices at each step of their reasoning process. For example, CoT-SC (Self-consistent CoT) \citep{wang2022self} uses the CoT approach to produce multiple reasoning paths and answers for a complex problem, selecting the most frequently occurring answer as the final output. Tree of Thoughts (ToT) \citep{yao2023tree} assumes that humans tend to think in a tree-like structure when making decisions on complex problems for planning purposes, where each tree node is a thinking state. It uses LLM to generate evaluations or votes of thoughts, which can be searched using breadth first search (BFS) or depth first search (DFS). These methods improve the performance of LLMs on complex reasoning tasks. DFSDT (Depth-first search-based decision tree) \citep{qin2023toolllm} employs a tree structure with each node representing a state that includes instruction context, prior API calls, and observations. The model not only reasons and moves to child nodes based on API calls but can also backtrack to explore alternative paths, ensuring a more diversified search and preventing cumulative errors. Graph of Thoughts (GoT) \citep{besta2023graph} further extends the reasoning paths from tree-structure to graph-structure, representing data produced by an LLM in the form of a flexible graph with individual units of information as nodes.

\end{itemize}

\cite{valmeekam2022large} concluded that ``LLMs are still far from achieving acceptable performance on common planning/reasoning tasks which pose no issues for humans to do.'' Similarly, \cite{wei2023chainofthought} also acknowledged this limitation, noting that ``we qualify that although chain of thought emulates the thought processes of human reasoners, this does not answer whether the neural network is actually reasoning.'' As a result, extensive efforts are still needed from the community to enhance the reasoning and planning ability of large language models.

\subsubsection{Self-Improving}
Just as kernel updates are driven by human feedback, focusing on bug reports and performance issues, leading to improvements in functionality, security, and stability, LLMOS also requires continuous enhancement to elevate its performance. This process involves the LLMs learning from interactions, refining their algorithms based on user experiences and feedback. By doing so, LLMs can develop new capabilities and skills, adapting to changing requirements and expectations. This iterative process of improvement ensures that LLMs remain effective, relevant, and efficient in handling diverse tasks and queries, mirroring the evolving nature of technology and user needs. After the LLM has been pre-trained, the learning strategies of LLM can be broadly categorized into two main types, as summarized and exemplified in the following.


\begin{itemize}
    \item \textbf{Learning from Feedback.} LLMs can learn from feedback about the consequences of its actions to the environment. For example, Reflexion, as proposed by \cite{shinn2023reflexion}, is a framework to reinforce language agents through linguistic task feedback rather than update weights, enabling them to learn from prior failings. Similarly, Recursively Criticize and Improve (RCI) \citep{kim2023language} is a prompting approach, which involves prompting the LLMs to identify issues in their output and subsequently enhance it based on the identified problems. Furthermore, these learning processes can be framed within the paradigm of Reinforcement Learning (RL). In this context, the LLM is trained to select actions that maximize a reward signal, aligning with the principles of RL. For example, \cite{ouyang2022training} presents Reinforcement Learning from Human Feedback (RLHF) to align LLMs with the feedback from human users; OpenAGI \citep{ge2023openagi} presents Reinforcement Learning from Task Feedback (RLTF), which learns from the performance of executing the LLM-generated task solution to fine-tune the planning strategy of the LLM.


    \item \textbf{Learning from Examples.} Recently, there has been a surge of interest in fine-tuning LLMs to perform tasks in a supervised way. For example, ToolLLaMA \citep{qin2023toolllm} is created by collecting various real-world APIs and generating instructions for their practical use, with solutions annotated using a language model such as ChatGPT and an efficient Depth-First Search Decision Tree. This process results in a dataset of instruction-solution pairs, on which a large language model such as LLaMA is fine-tuned to execute APIs according to instructions.
    Moreover, Gorilla \citep{patil2023gorilla} is trained by extracting key fields from API documentation and using GPT-4 to create instruction-API pairs, which are then used in a conversational format to fine-tune a model such as LLaMA, incorporating both retriever-aware and non-retriever training methods.

\end{itemize}




\subsection{Context Window (as Memory)}


In LLMOS, memory functions similarly to the context window in an LLM, defining the scope of information that the LLM can utilize and learn from while producing outputs. Increasing the amount of memory is desirable but always at a high cost.

Within the framework of an LLM, an expansion of the context window precipitates increased computational demands. This escalation is attributed to the quadratic computational complexity that is a characteristic of the attention mechanism employed in these models \citep{vaswani2017attention}. During the training phase, one widely adopted strategy to avoid the exorbitant costs associated with lengthy context is to conduct an initial pre-training phase using a relatively limited context window, and subsequently followed by a fine-tuning stage employing an expanded context window \citep{chen2023extending}. Consequently, two primary challenges emerge: 1) the reduction of computational costs associated with processing long contexts, and 2) the development of a flexible position encoding mechanism suitable for extended contexts.

\begin{itemize}
    \item \textbf{Efficient attention}: Various methods have been proposed to reduce the complexity of the multi-head attention mechanism, which can be categorized into three primary types: 1) Sparse attention mechanism \citep{zaheer2020big, gray2017gpu, kitaev2020reformer, beltagy2020longformer, ainslie2020etc, wang2020linformer} redefines the traditional computation of attention weights. In this approach, each query tensor is limited to engaging with only a subset of key tensors, as opposed to the entire set. This method effectively zeroes out other attention weights, thereby diluting the computation and reducing the overall computational burden; 2) Linear attention mechanism \citep{katharopoulos2020transformers}, which modifies the tensor multiplication process to be linear with respect to the sequence length without reducing the interaction between query tensors and key tensors; 3) Traditional multi-head attention uses multiple heads to split the query, key, and value tensor. Another method of reducing complexity aims at sharing heads for keys and values to optimize memory usage. By sharing heads between the keys and values in the multi-head attention setup \citep{shazeer2019fast, ainslie2023gqa}, it reduces the storage requirements for the key-value (KV) cache and improves efficiency.
    
    \item \textbf{Position encoding}: Position encoding can be classified into absolute position encoding and relative position encoding. The original Transformers paper \citep{vaswani2017attention} utilizes absolute position encoding, in which information is encoded in a combination of sin/cos functions whose wavelength increases from low- to higher-order dimensions of the embedding. Various methods, such as RoPE \citep{su2021roformer}, are proposed later to emphasize the relative position information between tokens. To extend the context window size, Alibi \citep{sun2022length}, LeX \citep{sun2022length}, and XPos \citep{press2021train} are proposed to enable length extrapolation so that a model can conduct inference on long context while being trained only on short context. To increase context length, methods such as position interpolation \citep{chen2023extending} and NTK-aware \citep{rozière2023code} position interpolation based on RoPE can be used.
\end{itemize}

Even though technically, the long context can be processed, there is no guarantee that the information in the long context can be correctly used and learned by LLM. Recent research has found that LLM does not robustly make use of information in long input contexts \citep{liu2023lost,shi2023large}. In particular, performance is often highest when relevant information occurs at the beginning or end of the input context and significantly degrades when models must access relevant information in the middle of long contexts. Similar findings are also found in chat-based LLM \citep{yang2023can}, where model's ability to track and enumerate environment states is unsatisfying. Given long documents, various prompt compression methods, such as LLMLingua \citep{jiang2023llmlingua}, have been proposed to reduce context length by removing unimportant tokens from the text.

\subsection{External Storage (as Files)}
In addition to generating content based on knowledge acquired during pre-training, LLMs also utilize externally stored information for several purposes. These include enhancing predictions in domain-specific areas, generating more current information based on recent updates, and improving long-term memory retention. The external storage functions similarly to a file system within an operating system and supports various data formats. Certain formats enable the LLMs to swiftly query data as required, while others act as auxiliary knowledge bases. These knowledge are accessible for instant access to provide information in response to queries requiring high precision or expert knowledge. This section will commence with an explanation of how data is stored, followed by a discussion on the methodologies employed to retrieve pertinent information from the data storage.
 
\subsubsection{Data Formats}
A file system in OS stores, organizes, and retrieves information in different modalities, including but not limited to plain text, images, audios, or videos. Similarly, LLM stores and retrieves data in different formats, such as natural language, embedding vectors, or graphs. It should be emphasized that these formats are not mutually exclusive, and many models incorporate multiple formats to concurrently harness their respective benefits. In the following, we introduce several representative data formats, each with its distinct features and applications:

\begin{itemize}
    
    \item \textbf{Embedding Vectors.} Embedding vectors represent words, phrases, or even entire documents as dense vectors in high-dimensional spaces. This format is crucial for processing information in machine-readable forms, such as natural language and images, enabling LLMs to efficiently and accurately retrieve necessary information for content generation. We note that dense vector retrieval, as exemplified by \cite{karpukhin2020dense}, generally serves as a necessary intermediate step for various data formats. Many different data formats are represented as dense vectors during this retrieval process. Specifically, when introducing embedding vector-based data formats, we refer to methods that explicitly store information in vector format. This approach is commonly used in conversational agents to maintain long-term memories. For example, \cite{zhong2023memorybank} identified the lack of long-term memory as a significant limitation in current LLM-based applications such as ChatGPT. This issue is particularly noticeable in scenarios that require sustained interactions, such as personal companionship, psychological counseling, and secretarial tasks. To equip LLMs with the ability to effectively access long-term memory, \cite{zhong2023memorybank} proposed ``MemoryBank,'' a novel vector-retrieval mechanism designed specifically for LLM-based applications. MemoryBank stores user-system past interactions, such as dialogues, as time-aware dense vectors known as memory pieces, with a dual-tower dense retriever and a memory updater inspired by Ebbinghaus' Forgetting Curve theory. A chatbot powered by MemoryBank demonstrates the effectiveness of this mechanism in long-term conversation. Similarly, \cite{zhao2023unimc} stores long-term conversational data as vectors for open-domain dialogue applications, where each piece of dialogue or summarized memory is transformed into a high-dimensional vector that captures its semantic meaning. Contrastive representation learning is used to increase the accuracy of memory retrieval. \cite{lee2023prompted} proposed a memory-enhanced chatbot to achieve long-term consistency and flexibility in open-domain conversations. It uses a summarizer model to summarize dialogues and store them as dense vectors in a memory pool after a certain number of rounds. Then, the relevant memory is retrieved to help generate responses. Later, \cite{lu2023memochat} introduced ``Memochat,'' which proposed an iterative ``memorization-retrieval-response'' cycle to maintain consistency in long conversations covering multiple topics. In this cycle, LLMs are trained to create structured memos, which help to keep track of the conversation context and topics.

    \item \textbf{Plain-Text Documents.} Language models that generate responses by retrieving external plain-text documents can greatly improve the quality of responses. This is especially true for tasks that demand domain-specific knowledge or up-to-date information, such as question answering \citep{guu2020retrieval, borgeaud2022improving} and fact verification \citep{lewis2020retrieval, izacard2022few}. Plain-text documents, which consist of unstructured text data, are often vast in size and contain a wealth of information. Examples of such retrieval include Wikipedia articles \citep{guu2020retrieval}, textbooks \citep{wang2023augmenting}, and programming code \citep{zhou2022docprompting}. Retrieving information from these documents poses a challenge due to their dense and extensive nature. Therefore, advanced retrieval methods such as Dense Passage Retrieval (DPR) \citep{karpukhin2020dense} are frequently utilized in the retrieval process. Retrieval-augmented LLM based on external document collections is an important extension of LLMs, where the system is not just reliant on pre-trained knowledge but also actively retrieves external knowledge for better response generation. The augmentation allows for more accurate, up-to-date, and context-relevant responses, particularly crucial for tasks involving real-time data or specialized knowledge. Research in this field is often referred to as Retrieval-Augmented Generation (RAG), focusing on critical issues such as pre-training language models in retrieval contexts and fine-tuning them for downstream tasks.
    
    \item \textbf{Structured Data.}
    Structured data is a commonly used format for storing external, useful knowledge for LLMs. Since much information naturally exists in structural formats, this enables models to access a wide range of information within complex knowledge structures. The sources of structured data are diverse, with knowledge graphs being one of the most pertinent formats for augmenting LLMs' generative abilities: \cite{pan2023unifying} explicitly discuss the use of knowledge graphs to address the hallucination issues in LLMs and enhance their interpretability. Furthermore, several popular LLM tools, such as LlamaIndex \citep{Liu_LlamaIndex_2022}, offer options to leverage existing knowledge graphs to improve knowledge generation. Another potentially valuable data format is tabular data. As introduced in \cite{sundar2023ctbl}, a novel concept of Conversational Tables (cTBLS) is designed to enhance the capabilities of conversational AI by integrating tabular data. This work utilizes a dense table retrieval method to rank relevant table cells. Subsequently, the responses of LLMs are grounded in the tabular information and the conversational context.

\end{itemize}





\subsubsection{Data Retrieval Methods}
\label{sec:rag}

The ability to access relevant and accurate information from external storage through sophisticated data retrieval methods is crucial for LLMOS's execution of targeted actions. A primary challenge in this process is selecting the most appropriate data file from an extensive repository of information. These data retrieval methods operate automatically, leveraging advanced algorithms and machine learning techniques.

For instance, \cite{zhu2023ghost} show that LLM can store past accomplished sub-goals of video games using a dictionary, where the sub-goals are keys, and the corresponding action sequences are values. When encountering familiar objectives, the first action is easily retrieved using the name of the goal. Conversely, \cite{park2023generative} propose a more sophisticated ``memory stream'' to record agents' past experiences in a list, labeled with text descriptions and timestamps of creation or last interaction. This data storage strategy enables agents to effectively retrieve useful experiences based on their current situation, using scores of Recency, Importance, and Relevance. Similarly, \cite{zhong2023memorybank} discuss storing detailed records of daily conversations, summaries of past events, and assessments of user personalities as vector representations indexed by FAISS \citep{johnson2019billion}, a library used for efficient similarity search in stored vectors. Furthermore, \cite{hu2023chatdb} highlight the limitations of conventional neural memory mechanisms, which are not symbolic and rely on vector similarity calculations, being prone to errors. It suggests the use of databases as an external symbolic memory for LLMs. Complex problems are decomposed into a series of SQL-based memory operation steps, greatly simplifying the retrieval process.

On the other hand, retrieving information from extremely large external documents heavily relies on existing methods in Information Retrieval (IR) research \citep{croft2010search}. Modern Information Retrieval involves two key stages: retrieval and ranking. The retrieval stage focuses on fetching relevant documents based on user queries using algorithms such as vector space models \citep{salton1975vector, robertson2009probabilistic}, or pre-trained models such as BERT \citep{devlin2018bert, karpukhin2020dense}. LLM extensively applies these methods for applications that depend on generating comprehensive domain knowledge or accurate information. To ensure better accuracy in retrieving information from a vast amount of documents, effective index representations are usually learned during pre-training or fine-tuning \citep{guu2020retrieval, borgeaud2022improving, lewis2020retrieval, izacard2022few, hua2023how}.




\subsection{Tools (as Devices/Libraries)}


In a traditional OS, peripheral devices, such as keyboards, mice, and printers, extend the system's capabilities, allowing for diverse forms of input and output that enhance the overall functionality of the computer. There are also various programming libraries, which include a diverse set of reusable functionalities that can be leveraged by applications through API calls.
Similarly, in the context of AIOS, hardware tools can be seen as analogous to these peripheral devices, and software tools can be seen as analogous to these libraries and APIs. 
These tools can range from data analysis modules to interactive interfaces, each adding a unique dimension to the LLM's processing and response abilities. They allow the LLM to interact with different data types, environments, and user requirements, significantly enriching its functionality. 
As a result, these hardware and software tools help the LLM to interact with the physical and digital worlds, expanding LLM's capabilities, and can be leveraged by agents for complex task solving. 
In the upcoming sections, we will delve into the specifics of these tools, illustrating how they enrich the LLM's capacity within the AIOS.

\subsubsection{Tool Categories}

Though LLMs are adept at handling numerous tasks, they encounter limitations in complex tasks that require deep domain knowledge or interaction with the external world. External tools enable LLMs to harness various resources and specialized knowledge, bolstering their capabilities. In the following, we present several representative tools for LLMs, as discussed in the existing literature.

\begin{itemize}
    \item \textbf{Software Tools} are domain expert models or APIs that help LLMs to finish a specialized sub-task, such as searching a query on the Web through a search API or checking the weather through a third party weather service API. 
    Recent trends show a growing integration of APIs with LLMs, serving as interfaces for external programs to interact with LLMs, and acting as a bridge between the LLM and other software applications, thus extending LLMs' capabilities across various applications and services. For instance, OpenAGI \citep{ge2023openagi} trains LLMs to use various domain expert models as tools for reasoning, planing, and complex task solving based on reinforcement learning from task feedback (RLTF). TPTU \citep{ruan2023tptu} interfaces with both Python interpreters and LaTeX compilers for mathematical computations. Gorilla \citep{patil2023gorilla}, a fine-tuned LLM, is engineered to generate precise API call arguments and prevent hallucinations. ToolLLM \citep{qin2023toolllm} presents a general framework for tool use, including data construction, model training, and evaluation. It also provides an instruction-tuning dataset for tools, collected from over 16,000 real-world APIs. TaskMatrix.AI \citep{liang2023taskmatrix} connects foundational models with millions of APIs for diverse task completion, facilitating user interaction in an interactive manner.
    ChemCrow \citep{bran2023chemcrow} integrates several expert-designed models to augment LLMs in chemistry-related tasks such as organic synthesis, drug discovery, and materials design. MM-REACT \citep{yang2023mm} combines LLMs with various vision expert models for advanced multi-modal reasoning and actions. Using expert models as tools, LLM agents can tackle advanced tasks necessitating expert knowledge.
    
    \item \textbf{Hardware Tools.} While the aforementioned tools enhance LLMs in the digital world, physical tools such as robotics and embodied AI serve as pivotal means to connect LLMs with the physical world. These tools enable LLMs to actively observe, understand, and interact with their physical surroundings. Observations allow LLMs to gather various inputs from the physical world, converting them into multi-modal signals to augment actions such as navigation and manipulation. For example, Soundspace \citep{chen2020soundspaces} explores observing physical space geometry using reverberating audio sensory inputs. Physical tools enable LLMs to execute user commands, transcending the role of merely providing natural language instructions. SayCan \citep{ahn2022can}, for instance, incorporates physical tools into LLMs for real-world tasks such as cleaning tabletops or retrieving items. In essence, physical tools act as the ``hands, ears, eyes'' etc. of LLMs to interact with the physical world, fostering real-world grounding.

    \item \textbf{Self-made Tools.} Current tools are primarily designed by humans. Recently, there are increasing interest in the use of LLMs for automated tool making. This involves generating executable programs or enhancing existing tools to create more powerful solutions, guided by appropriate instructions and demonstrations \citep{qin2023tool, qian2023creator, chen2021evaluating}. For example, a large code model, as evaluated in \cite{chen2021evaluating}, is capable of generating executable programs based on user-provided instructions. These programs can then serve as specialized tools to address particular tasks. Furthermore, these LLMs can also acquire the ability to self-debug, which is an essential skill for maintaining and improving the tool functionality, as detailed in \cite{chen2023teaching}.
    
\end{itemize}


\subsubsection{Tool-Driver and Tool-API}

In traditional OS, Drivers or APIs play a pivotal role in enabling the system to interact with specific Devices or Libraries. The drivers provide interfaces to connect the OS with hardware devices, while the APIs provide interfaces to connect the OS or application with software libraries.
In the context AIOS, where hardware tools are viewed as devices and software tools are viewed as libraries, Tool-Drivers and Tool-APIs are required, which serve as the interface for AIOS or agents to use these hardware and sofware tools, respectively. 

Existing literature usually defines the Tool-Drivers and Tool-APIs in the form of prompts. Specifically, these prompts are composed of two essential elements: application scenarios and invocation methods. Much like how Drivers or APIs in a traditional OS control access to specific Devices or Libraries, commonly used prompts in AIOS should equip LLMs with an in-depth understanding of application scenarios. This enables LLMs to judiciously determine the appropriateness of a particular tool for a given task. Moreover, in parallel to how Drivers or APIs facilitate the communication between the OS and devices, tool instruction prompts should clearly outline the invocation methods. This is crucial for LLMs to comprehend the inputs and outputs of the tools, ensuring their effective execution and integration into the system.

Utilizing the inherent zero-shot and few-shot learning capabilities of LLMs \citep{radford2019language,brown2020language}, agents can gain insights about tools through zero-shot prompts that elucidate tool functionalities and parameters, or few-shot prompts offering demonstrations of particular tool usage scenarios and methodologies \citep{schick2023toolformer,parisi2022talm}. Usually, a single tool is inadequate for complex tasks. Hence, agents must adeptly decompose these tasks into manageable subtasks, where their understanding of the tools is pivotal. After understanding individual tools, LLMs should determine their application in addressing complex tasks. One approach is to generate actions by extracting pertinent information from memory relevant to the current task. For instance, Generative Agents \citep{park2023generative} maintain a memory stream, consulting it for recent, pertinent, and crucial information before each action to guide their decisions. In GITM (Ghost in the Minecraft) \citep{zhu2023ghost}, to achieve specific sub-goals, the agent probes its memory to identify if any similar tasks have been successfully executed before. If so, it replicates those successful actions for the current task. In collaborative frameworks such as ChatDev \citep{qian2023communicative} and MetaGPT \citep{hong2023metagpt}, agents engage in mutual communication, retaining the conversation history in their memories. Another strategy involves executing a pre-formulated plan. For example, in DEPS (Describe, Explain, Plan and Select) \citep{wang2023describe}, given a task, if there are no indicators of the plan's failure, the agent proceeds with actions based on that plan.


%





\section{AIOS-Agent Ecosystem} \label{sec:ecosystem}

\subsection{Agents as Applications}

On top of the LLMOS-based AIOS, a diverse scope of AI Agent applications can be developed, akin to traditional OS-based applications such as browsers and photo-editing softwares. These Agent Applications (AAPs) generally comprise three components: the agent profile, an accessible external database, and task-specific tools.
On one hand, the agent profiles are written into the prompt and used to indicate the agent functionality, influencing the LLM behaviors to exhibit certain roles such as a coder agent, a teacher agent, or a travel planning agent, as described in sources such as \cite{qian2023communicative, li2023camel, ge2023openagi}. Typical agent profiles may encompass basic information such as age, gender, and career \citep{park2023generative}, or psychology information reflecting the personalities of the agents \citep{serapiogarcía2023personality}, as well as social information detailing the relationships between agents \citep{li2023camel}. The nature of the agent's profile is tailored to the application's context; for example, applications focused on human cognitive processes will emphasize psychological information.
Furthermore, the inclusion of an external database and toolset within the prompt enhances the LLM's interaction with the external world, specifying elements such as file locations and tool descriptions.

\begin{figure}[h]
\centering
\includegraphics[width=\textwidth]{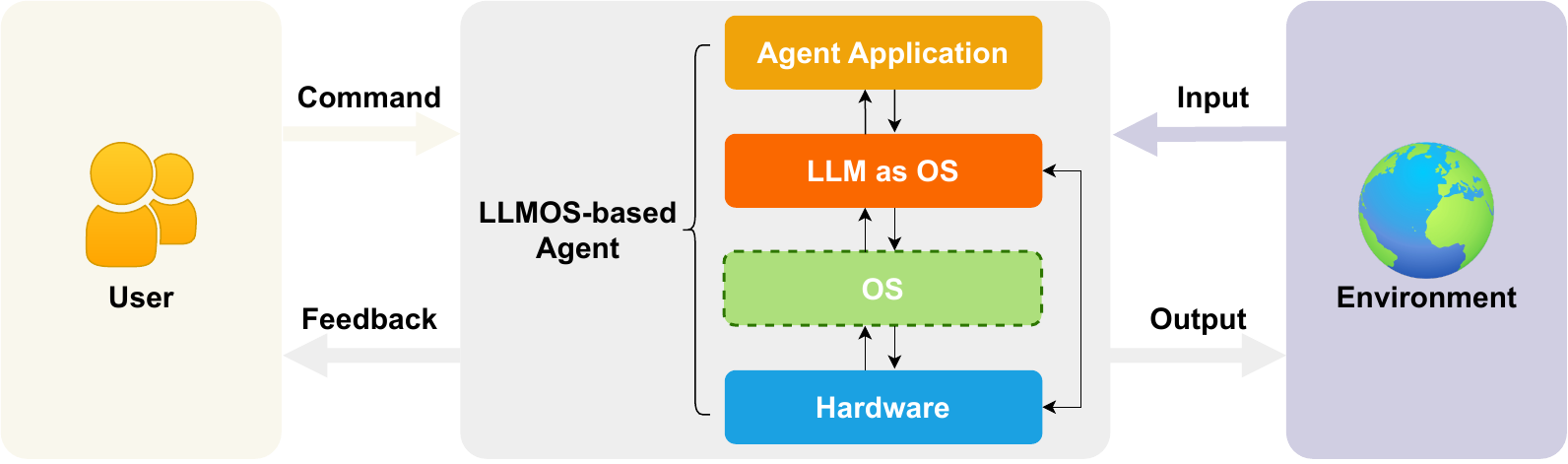}
\caption{An illustration of LLMOS-based AI Agent.}
\label{fig: agent}
\end{figure}

As illustrated in Figure \ref{fig: agent}, by merging the LLMOS layer with the OS layer, Hardware layer, and the Agent Application layer, we can establish an autonomous LLMOS-based Agent system. This system will respond to natural language instructions from users, capable of executing a variety of tasks through its interactions with the environment and its inherent knowledge.

\subsection{Natural Language Programming for Agents}\label{sec:nlprog}

Natural Language Programming (NLProg) acts as a crucial intermediary between LLMOS and Agents within the AIOS-Agent ecosystem. This innovative approach allows average users to easily program Agent Applications (AAPs) using natural language. This development democratizes the creation and accessibility of computer software, representing a significant shift from the traditional OS-APP ecosystem. Traditionally, desktop or mobile applications (APPs) required programming by skilled software developers using professional programming languages. In contrast, NLProg empowers even those without formal training in these languages to develop applications.

This trend aligns with the historical evolution of programming languages, which have progressively become more user-friendly. The journey from binary codes to assembly language and then to various high-level languages such as C, C++, Java, and Python reflects this trend. The introduction of Natural Language as a Programming Interface is a natural progression in this evolution. It simplifies programming for the average user, allowing them to program agent applications and interact with computers without the need for specialized training in conventional programming languages.

Moreover, this shift has broader implications for the field of computer science and technology. It opens up new possibilities for innovation and creativity, as a wider range of individuals can contribute to software development. This inclusivity can lead to the development of more diverse and tailored applications, as people from different backgrounds and with varying expertise can bring their unique perspectives to software design. Additionally, NLProg in the AIOS ecosystem fosters a more intuitive interaction between humans and computers, enhancing user experience and potentially leading to more efficient and effective use of technology in various sectors. As this approach gains traction, it could significantly alter the landscape of technology development, making it more accessible and aligned with the natural human way of communication and understanding.

\subsection{The Ecosystem}
In the AIOS-Agent ecosystem, the potential for collaboration and networking among agents heralds a new era of digital assistance and decision-making. These agents, with their diverse skill sets and role profiles, can work in tandem to address complex challenges, offering a multifaceted approach to problem-solving. For instance, an agent with a background in data analysis can collaborate with another agent specialized in creative design, leading to innovative solutions that a single agent might not conceive.

Moreover, the AIOS ecosystem is designed to be scalable, accommodating an increasing number of agents as user needs grow and evolve. This scalability ensures that the system remains efficient and effective, even as the complexity of tasks increases. The AIOS also emphasizes the importance of learning and adaptation. Agents in this ecosystem can learn from their interactions, both with users and other agents, continuously evolving and enhancing their capabilities. This feature is crucial for keeping up with the rapidly changing landscape of technology and user expectations.

In the broader context, the AIOS-Agent ecosystem can significantly impact various industries, from healthcare, where agents can assist in patient care and medical research, to education, where agents can offer personalized learning experiences. The versatility and adaptability of this ecosystem make it a valuable asset in any field where decision making, data analysis, and creative problem solving are crucial. As this technology matures, it holds the promise of transforming our interaction with the digital world, making it more intuitive, efficient, and responsive to our needs.

\section{LLMOS in Practice: AI Agents} \label{sec:agents}

In this section, we present a comprehensive overview of the current applications of LLMOS-based agents, aiming to provide a wide-ranging perspective on their practical deployment scenarios.
Specifically,  we mainly introduce three scenarios: single agent applications, multi-agent applications, and human-agent applications.

\subsection{Single Agent Applications} \label{sec:single}


Single agent applications mainly focus on utilizing a single agent to address various user tasks. We categorize single agent applications into two distinct types based on the external environments they interacted with, i.e., physical environment and virtual/digital environment.

\subsubsection{Physical Environment}


Unlike virtual or simulated environments, the physical environment is made up of tangible elements that an AI Agent must navigate or manipulate. This concept is particularly relevant in the field of robotics and embodied AI, where the agents are not just software algorithms but have a physical presence or are integrated with physical systems. Following this, we present a range of exemplary scenarios, as detailed in existing literature.

\begin{itemize}
    \item \textbf{Scientific Research.} Agents have the capacity to function autonomously, undertaking experiments independently, while also serving as invaluable resources for scientists engaged in research projects \citep{boiko2023emergent, bran2023chemcrow}. For instance, \cite{boiko2023emergent} propose an end-to-end platform that automates scientific experimentation, integrating AI modules, reasoning capabilities, software tools, and laboratory hardware. It autonomously performs tasks ranging from online information gathering and experiment design to running protocols and controlling robotic equipment, while adapting to errors and refusing unethical requests. ChemCrow \citep{bran2023chemcrow} is a suite of 17 specialized tools designed to aid chemical research, offering methodological suggestions and highlighting safety hazards based on the input objectives. \cite{huang2023benchmarking} also propose MLAgentBench, a suite of ML tasks for benchmarking AI research agents.

    \item \textbf{Robotics.} Recent studies have employed LLM-based agents in the fields of Robotics. For example, ChatGPT for Robotics \citep{vemprala2023chatgpt} employs ChatGPT for a wide array of robotics tasks through strategic prompt engineering, showing its ability to comprehend and respond to natural language instructions in the context of robotics applications. SayCan \citep{ahn2022can} comprises two integral components: the ``Say'' part of the LLM, responsible for task-grounding by identifying pertinent actions for a high-level objective, and the ``Can'' part, which encompasses the learned affordance functions that offer a world-grounding, thereby determining the feasible actions for the plan's execution. 
It ensures not only the relevance of the actions chosen for the specified task but also their feasibility in the real-world scenario.
    VOYAGER \citep{wang2023voyager} presents lifelong learning with prompting mechanisms, skill library, and self-verification, which are based on LLM. These three modules aim to enhance the development of more complex behaviors of agents. 



 \item \textbf{Autonomous Driving.} Recent studies have harnessed LLMs to enhance self-driving car technologies. For example, \citet{fu2023drive} propose an agent-based autonomous driving system, which widely adopts a four-module framework: Environment, Agent, Memory, and Expert. Within this framework, the Environment sets the interaction stage, the Agent perceives the environment and makes decision, the Memory accumulates experience for action, and the Expert provides training advice and inconsistency feedback. 
 
\end{itemize}


\subsubsection{Virtual/Digital Environment}
Agents in virtual or digital environment mainly includes the manipulation of APIs \citep{schick2023toolformer,yao2023impact,ge2023openagi,parisi2022talm,tang2023toolalpaca}, accessing the Internet and websites \citep{nakano2022webgpt}, executing codes \citep{zhang2023toolcoder}, and simulation in historical settings \citep{hua2023war}. Such digital grounding is cheaper and faster than physical or human interaction. It is thus a convenient test bed for language agents and has been studied with increasing intensity in recent years. 
In the following, we present several representative scenarios as studied in the existing literature.

\begin{itemize}

    \item \textbf{Coding.} This category focuses on leveraging the capabilities of agents to generate programs. For example, ToolCoder \citep{zhang2023toolcoder} is a system that merges API search tools with existing models to facilitate code generation and API selection, using a two-step approach. Initially, an automated data annotation technique involving ChatGPT embeds tool usage data into the source code, followed by fine-tuning the code generation model; during inference, the API search tool is integrated to autonomously suggest API choices, optimizing code generation and improving API selection decision-making. Moreover, Lemur-series models \citep{xu2023lemur} are meticulously pre-trained and instruction fine-tuned to demonstrate balanced language and coding capabilities.

    \item \textbf{Web Service.} This category primarily revolves around utilizing agents to address web-based tasks through diverse APIs. For example, Auto-GPT \citep{Significant_Gravitas_Auto-GPT} is an automated agent designed to set multiple objectives, break them down into relevant tasks, and iterate on these tasks until the objectives are achieved. OpenAGI \citep{ge2023openagi} is an LLM-based agent designed for reasoning, planing, and executing tools to achieve complex tasks, accompanied with a benchmark to evalute the agent's task-solving performance.
    BMTools \citep{qin2023tool} is an open-source repository that extends LLMs with tools and provides a platform for community-driven tool building and sharing. It supports various types of tools, enables simultaneous task execution using multiple tools, and offers a simple interface for loading plugins via URLs, fostering easy development and contribution to the BMTools ecosystem. 
    Mind2Web \citep{deng2023mind2web} provides a benchmark for developing and evaluating generalist agents for the Web, which are agents that can follow language instructions to complete complex tasks on websites.
    MusicAgent \citep{yu2023musicagent} integrates numerous music-related tools and an autonomous workflow to address user requirements. Auto-UI \citep{zhan2023you} is a multi-modal solution that directly interacts with the interface, bypassing the need for environment parsing or the dependence on application-dependent APIs.


\item \textbf{Games.} This includes agents interacting in the game environments \citep{hausknecht2020interactive,cote2019textworld,shridhar2020alfworld}. For example, MineClip, introduced by \cite{fan2022minedojo}, is a novel agent learning algorithm that leverages large pre-trained video-language models as a learned reward function. Based on it, the authors further proposed MineDojo, a framework built on the popular Minecraft game that features a simulation suite with thousands of diverse open-ended tasks and an internet-scale knowledge base with Minecraft videos, tutorials, wiki pages, and forum discussions. 

\item \textbf{Recommendation.} Recent literature demonstrates the efficacy of employing LLM and Agents in recommender systems \citep{geng2022recommendation, wang2023recmind, feng2023large, wang2023recagent}. For instance, RecMind \citep{wang2023recmind} develop an LLM-based recommender system agent, which provides personalized recommendations based on planning, use of tools, obtaining external knowledge, and leveraging individual user's personalized data. LLMCRS \cite{feng2023large} is a conversational recommendation agent that utilizes Large Language Models (LLMs) for efficient sub-task management during the recommendation process. It combines LLMs with expert models for specific sub-tasks and employs LLMs as a language interface for generating improved user responses, thereby enhancing overall performance and response quality in conversational recommendation systems.

\end{itemize}

\subsection{Multi-Agent Applications} \label{sec:multi-agent}




Multi-Agent Systems (MAS) \citep{wooldridge1995intelligent}  emphasize the coordination and collaboration among a group of agents to effectively solve problems. The existing LLM-based MAS landscape is broadly categorized into two types: Collaborative Interaction and Adversarial Interaction. 


\subsubsection{Collaborative Interaction} 
As the scope and complexity of tasks amenable to Large Language Models (LLMs) increase, a logical strategy to augment the effectiveness of these agents is to employ cooperative multi-agent systems. Such systems, prevalently utilized in practical applications, operate on the principle where each agent evaluates and understands the requirements and capabilities of its peers, thereby fostering a collaborative environment conducive to shared actions and information exchange \citep{li2023camel}.
In the specific area of Non-Player Characters (NPCs), the concept of Generative Agents \citep{park2023generative} emerges as a compelling simulation of human behavior within interactive applications. This approach is exemplified by the deployment of twenty-five agents in a sandbox environment akin to The Sims, allowing users to engage with and influence the agents as they execute daily routines, interact socially, establish relationships, and organize group activities. Furthermore, the Humanoid Agents system \citep{wang2023humanoid} enhances the realism of Generative Agents by incorporating three fundamental aspects of ``System 1'' processing: the fulfillment of basic needs (such as hunger, health, and energy), emotional responses, and the dynamics of interpersonal relationships.
In the field of Software Development, the MetaGPT system \citep{hong2023metagpt} represents a specialized LLM application that leverages a multi-agent conversational framework. This innovative framework facilitates automatic software development by assigning distinct roles to various GPT models, enabling them to collaborate effectively in the creation of software applications. Additionally, BOLAA \citep{liu2023bolaa} introduces a controller module that orchestrates the coordination and communication among multiple collaborative agents, thereby streamlining the selection and interaction processes between different labor agents. CHATDEV \citep{qian2023communicative} proposes an advanced software development framework that utilizes agents to foster enhanced collaboration among the diverse roles integral to the software development cycle.
In the domain of Conversational AI, research exemplified by \cite{fu2023improving} delves into the potential of LLMs to autonomously refine their negotiation skills. This is achieved through engaging the models in bargaining games against one another, complemented by the integration of natural language feedback from an AI critic. This study underscores the evolving capabilities of LLMs in complex, interactive settings.

\subsubsection{Adversarial Interaction}
Traditionally, collaborative methods have been extensively explored in multi-agent systems. However, researchers are increasingly recognizing that introducing concepts from game theory into systems can lead to more robust and efficient behaviors.
For example, \cite{du2023improving} introduce the concept of debate, endowing agents with responses from fellow peers. When these responses diverge from an agent's own judgments, a ``mental'' argumentation occurs, leading to refined solutions. ChatEval \citep{chan2023chateval} establishes a role-playing-based multi-agent referee team. Through self-initiated debates, agents evaluate the quality of text generated by LLMs, reaching a level of excellence comparable to human evaluators. Corex \citep{sun2023corex} is constituted by diverse collaboration paradigms including Debate, Review, and Retrieve modes, which collectively work towards enhancing the factuality, faithfulness, and reliability of the reasoning process. These paradigms foster task-agnostic approaches that enable LLMs to ``think outside the box,'' thereby overcoming hallucinations and providing better solutions. MAD (Multi-Agent Debate) \citep{liang2023encouraging} is a framework wherein several agents engage in a ``tit-for-tat'' exchange of arguments under the oversight of a judge who steers the discussion towards a conclusive solution. Furthermore, WarAgent \citep{hua2023war} considers each country as an LLM-based agent and simulates the international conflicts among the countries using World War I, World War II, and the Warring States Period in Ancient China as examples, which showcases possible approaches towards LLM multi-agent based policy simulation and answering the ``what if'' questions for historical analysis.








\subsection{Human-Agent Applications} \label{sec:human-agent}
Most existing agent frameworks often limit themselves to defining and controlling agent behavior through system prompts, allowing the agent to independently plan and act. A notable shortcoming of this approach is the restricted, and sometimes non-existent, capacity for meaningful interaction between human users and agents, including multi-agent setups.
Addressing this gap, AutoGen \citep{wu2023autogen} offers an open-source solution enabling developers to construct LLM applications through multiple agents. Specifically, AutoGen distinguishes itself with agents that are not only customizable and conversable, but also versatile in their operational modes, which incorporate a blend of LLMs, human inputs, and tools, enhancing the interaction capabilities and efficiency of the agents.
Furthermore, AGENTS \citep{zhou2023agents} introduces a novel approach to creating controllable agents. This method involves the use of symbolic plans or standard operating procedures (SOPs), which can be generated by an LLM and subsequently modified by the user. This feature allows for greater customization and fine-tuning of agents, providing a more user-centric and adaptable agent framework.
Collectively, these developments represent a significant shift in the landscape of agent frameworks, moving towards systems that not only automate tasks but also facilitate a more interactive and collaborative environment between humans and agents. 

\section{OS-inspired Future Directions}
\label{sec:future}

The evolution history of operating system over the past half century has witnessed the continuous development of computer hardware and the explosive growth of data, which empowers the fast iteration of Artificial Intelligence in the past decade. In this section, we enumerate the lessons learned from the history of operating systems and provide envisions of the future directions for AIOS.

\subsection{Resource Management}

\subsubsection{Memory Management}

Physical memory (DRAM) has always been an insufficient resource from the beginning age of computer systems until now. To reduce the tension between users' requirements and the fact of DRAM resource shortage, several approaches have been proposed in modern operating systems.

\begin{itemize}
\item \textbf{Swapping to external storage.} A dedicated partition in external storage is reserved for swapping unused memory regions to external storage in a user-transparent way, as detailed in~\cref{bg:os:kernel}. This approach enlarges the available DRAM resources in the system.
\item \textbf{Memory sharing.} To support data sharing across applications, modern operating systems provide memory sharing between applications, which provides both sharing and also reduces the extra copies of shared data.
\item \textbf{Memory disaggregation.} Entering the terabyte scale data, fitting the memory requirements for an application in a single operating system on a sole machine is becoming challenging. To remedy this, disaggregated memory techniques were proposed to allow an application to use memory on another machine via network. Moreover, recent Compute eXpress Link (CXL) technique \citep{CXL} significantly reduces the software overheads and hardware latency of remote memory access. 
\end{itemize}

LLMs have revolutionized abilities, but are constrained by limited context windows, hindering their utility in tasks such as extended conversations and document analysis.
To mitigate the limited context window problem in LLMs, approaches inspired by the above operating system memory management methods can be developed. A swapping mechanism, similar to OS swapping to external storage, can be implemented in LLMs to temporarily store inactive parts of their context, effectively enlarging the context window. This would require efficient retrieval systems to minimize latency. For example, MemGPT \citep{packer2023memgpt} intelligently manages different memory tiers in order to effectively provide extended context within the LLM's limited context window, and utilizes interrupts to manage control flow between itself and the user. Additionally, mirroring OS memory sharing, LLMOS can share context or learned patterns across different LLMOS-based Agent instances, reducing redundancy and allowing access to a larger shared knowledge pool for agents. Finally, drawing from memory disaggregation in modern OS, agents can leverage networked memory resources, enabling them to access and process data stored across multiple LLMOSes, thus expanding their abilities significantly. Each of these strategies, while offering potential solutions, also presents unique challenges such as managing latency, ensuring consistency in shared contexts, and handling the complexities of distributed memory systems.

\subsubsection{Tool Management}

Serving as external resources outside LLM, the hardware tools (e.g., robotics) and software tools (e.g., Web Search API) are the counterpart of devices and libraries in modern operating systems as shown in Table~\ref{tab:components}, respectively. Taking the example of the ecosystem in modern Linux operating system--the most widely-used open source operating system community by now, here are the successful experiences in its evolving history. Specifically, a rich set of built-in and third-party libraries from thousands of experts and open-source developers leads to the success of the Linux ecosystem. Managing the install/uninstall and dependencies of those libraries, along with the versioning for tracking software development, is critical. The Linux ecosystem, over the past few decades, provides library management tools such as \textit{dpkg}\footnote{\url{https://man7.org/linux/man-pages/man1/dpkg.1.html}} in Debian-based Linux distributions, and \textit{yum}\footnote{\url{http://yum.baseurl.org/}} in RHEL-based Linux distributions. \textit{Git}\footnote{\url{https://git-scm.com/}}, a distributed version control system that plays a crucial role in the development of the Linux ecosystem, allows collaborative and parallel development, boost the speed of development cycles in building the Linux ecosystem.

The process of code comparison in Git, particularly during merging and rebasing, typically analyzes changes at the line level rather than understanding the semantic meaning of the text. However, as discussed in Section \ref{sec:nlprog}, the development of agents using natural language is increasingly achievable. Adapting existing version control software, which is based on the code or texts without knowing the semantics and contexts, for use with natural languages poses distinct challenges. First, natural languages, while structured by grammars, exhibit a loosely coupled relationship with the varied expressions of different users. Second, this loose coupling can result in natural language statements that are semantically equivalent but differ in their wording. For example, in the context of an AI agent system, the instructions ``Analyze the latest sales data and generate a report'' and ``Generate a report based on the analysis of the most recent sales data'' convey the same instruction to the agent but are phrased differently. Incorporating the ability to recognize and reconcile these semantic nuances in natural language into version control software is essential. This is especially critical in collaborative development of complex AI agent systems, where such a feature can greatly streamline development cycles by effectively managing and merging diverse natural language inputs.

\subsection{Communication}


Domain-Specific Languages (DSLs) are widely used in both native operating systems and cloud environments, which address specific requirements or tasks within a particular domain. Operating systems are often equipped with scripting languages on top of the command-line interfaces that allow users to perform various tasks collaboratively. Unix-like operating systems use shell scripting languages such as \textit{Bash}\footnote{\url{https://www.gnu.org/software/bash/}}, which is designed for automating tasks in a command-line environment. In the cloud computing, DSLs are commonly used to define infrastructure as code~\citep{DSL-cloud}. It allows users to describe and provision cloud infrastructure resources, including virtual machines, networks, and storage; It can also be used for scheduling tasks on infrastructure resources. DSLs strike a balance between the underlying OSes and users that they are readable for both.

Leveraging the wisdom gleaned from OSes, multi-AIOS communication can be enhanced by adopting structured communication protocols that are analogous to the function of DSLs in simplifying complex tasks. Just as DSLs provide a medium for users to interact effectively with the operating system and its resources, a similar specialized protocol can be established for LLMs to communicate with one another. This protocol would standardize interactions, allowing for the clear transmission of context, tasks, and goals between LLMs, thus facilitating a more coordinated and coherent multi-agent operation. It would ensure that despite the varied functionalities and knowledge bases of individual LLMs, there is a common language or method through which they can collaborate, share insights, and synchronize their learning processes. This approach mirrors the way operating systems manage resources and processes, ensuring harmonious and efficient functionality across various components of a system.


In current AI Agents, the task-solving plan is still represented by natural language in most cases. However, in the future, we can even develop DSLs or semi-structured natural language grammars for representing Agent's task-solving plans. This involves the following breakthroughs:

\begin{itemize}
\item Defining Basic Operations: 
This would standardize the basic operations, tools, or commands that an AI Agent understands and can execute for task-solving.

\item Sequence of Operations: Task-solving plans would then be sequences of these basic operations, making them more structured and potentially more efficient.
\end{itemize}

Using LLM to interpret users' natural language instructions into a DSL-composed plan brings several benefits, including 1) Improved Consistency: Standardized operations would lead to more predictable and consistent outcomes; 2) Easier to Interpret: A well-defined DSL makes it easier for AI Agents to interpret and execute plans; and 3) Scalability: With a standard DSL, it is easier to scale solutions across different platforms and applications.

However, achieving this goal also meets some important challenges that need future research attention: 1) Complexity of Natural Language: Natural language is inherently ambiguous and context-dependent, making it challenging to convert instructions into structured plans consistently; 2) Flexibility vs. Standardization: Striking a balance between the flexibility of natural language and the rigidity of a standardized DSL; 3) Interoperability: Ensuring the DSL works well with a wide range of tools and platforms; and 4) Adaptation and Learning: The system needs to continuously learn and adapt to new instructions, tools, and tasks.

Overall, the use of a large language model as an interpreter (LLM as Interpreter), which can translate natural language described plans to DSL described plans, can be an important direction to create task-solving plans, and to bridge LLM as OS (LLMOS) and Agent Applications (AAP). The development of a DSL for such plans could further streamline and standardize the process, though it would come with its own set of challenges. This approach has the potential to make complex task execution more accessible and efficient, paving the way for more advanced AI systems in the future.

\subsection{Security}

Security has been an important issue as the wide-spread of operating systems from labs to our daily life in the 1980s. The consequences of the operating system vulnerabilities can be roughly categorized as following.

\begin{itemize}
\item \textbf{Breaking down the system.} In its early stages, viruses like the \textit{Morris Worm}~\citep{Morris-Worm} were primarily created to compromise operating systems, serving as a demonstration of individual programmers' prowess in hacking. While these attacks did not result in direct financial losses, users faced potential harm through the compromise of personal data or critical workplace documents.

\item \textbf{Racketeering.} In later stages, vulnerabilities within operating systems became targets for illicit activities, including the exploitation of users for racketeering purposes. An example of this is the \textit{WannaCry Ransomware}\footnote{\url{https://nvd.nist.gov/vuln/detail/cve-2017-0143}}, which encrypts users' files and demands a ransom for their release. Additionally, banking trojans like \textit{Zeus}\footnote{\url{https://nvd.nist.gov/vuln/detail/cve-2010-0188}} exploit vulnerabilities by intercepting communication channels between users and banking systems, resulting in direct financial losses for the affected users.

\item \textbf{Stealing Resource.} Malicious software, such as \textit{Coinhive}\footnote{\url{https://krebsonsecurity.com/2018/03/who-and-what-is-coinhive/}}, is crafted to harness the computing power of other machines for cryptocurrency mining, including Bitcoin. While it may not inflict direct harm on operating systems, the substantial utilization of CPU and memory resources can significantly impair the overall system performance. In cloud environments, this slowdown has the potential to translate into financial losses, making it imperative to address such threats proactively.

\end{itemize}

Security vulnerabilities in operating systems are hard to prevent. The state-of-the-art approaches detect and capture those malware and virus at different levels.

\begin{itemize}

\item \textbf{Static Analysis.} This approach conducts code-level or binary-level analyses by examining the code or binary image of an application or part of the OS. It is often performed when a third-party application is published to the cloud, or when a file is downloaded to the file system in a user's operating system.


\item \textbf{Fuzzing.} Fuzzing (Fuzz Testing) involves the automated generation of a large number of random or semi-random inputs to a program to discover vulnerabilities, crashes, or unexpected behaviors.






\end{itemize}

Similarly, adversarial attacks have been the subject of extensive study in LLM research \citep{wei2023jailbroken,zou2023universal,qi2023visual, yang2023comprehensive}, representing a significant threat to the security of AIOS. Furthermore, research \citep{yang2023shadow, qi2023fine} has revealed that an aligned LLM can be broken using a very small dataset comprising only a few hundred data points. This vulnerability is not only a significant concern in terms of system integrity but also raises alarming implications for the safety of agents interacting with these systems. How to be robust and defend against such attacks has yet to be studied in the context of AIOS, LLMOS, and Agents.

Looking into the future, as agent applications in AIOS evolve to be programmed by natural language, the intricacy of scanning the code of such applications will increase due to the expansive and less structured nature of natural language compared to the constrained syntax of programming languages. This necessitates a robust and effective scanning tool for AIOS agents. Moreover, fuzzing can be seen as an initial step towards creating a Red-teaming dataset for LLMOS-based Agents. For instance, \cite{ruan2023identifying} have developed a multi-agent system that generates red-teaming scenarios for LLM-based agents, using GPT-4 to simulate adversarial environments within textual scenarios. The study provides a rich array of scenarios that serve as a valuable resource for future research into the alignment of LLM-based Agents.

\section{Conclusions}

This paper presents a novel vision for the future of computing within the AIOS-Agent ecosystem, where the LLM functions as the core of AIOS. This innovative approach marks a significant departure from the conventional OS-APP ecosystem, heralding a new era in technology where AI and traditional computing systems merge seamlessly. The AIOS-Agent ecosystem envisaged here is not just an incremental change but a fundamental shift in how we interact with technology. By positioning LLM at the system level, Agents as applications, Tools as devices/libraries, and Natural Language as the Programming Interface, we redefine the interaction between users, developers, and the digital world. This paradigm shift promises to democratize software development and access, allowing users and developers to program Agent Applications (AAPs) using natural language. This accessibility contrasts sharply with the traditional ecosystem, where software development is confined to those with specialized programming skills. Moreover, the discussion of single and multi-agent systems, as well as human-agent interactions, illustrates the potential of AIOS in enhancing productivity, creativity, and decision-making processes across various domains. Looking ahead, the proposed strategic roadmap, informed by the developmental trajectory of the traditional OS-APP ecosystem, offers a pragmatic and systematic approach to the evolution of AIOS and its Agent Applications. This roadmap not only guides future development and research in this field but also anticipates the challenges and opportunities that lie ahead.

\newpage
\bibliographystyle{ACM-Reference-Format}
\bibliography{main}


\end{document}